\documentclass[sigconf]{acmart}
\usepackage{enumitem}
\usepackage{multirow}
\usepackage{todonotes}
\usepackage{color}
\usepackage{booktabs}

\setcopyright{acmcopyright}

\acmConference[JCDL '17]{The ACM/IEEE-CS Joint Conference on Digital Libraries}{June 2017}{Toronto, Ontario, Canada}
\copyrightyear{2017}
\acmPrice{15.00}
\acmISBN{123-4567-24-567/08/06}
\acmDOI{10.475/1234}

\begin{document}
\title{Understanding the Impact of Early Citers\\ on Long-Term Scientific Impact}

\author{Mayank Singh}
\affiliation{%
  \institution{Dept.  of Computer Science and Engg.}
  \streetaddress{IIT Kharagpur, India}
  }
\email{mayank.singh@cse.iitkgp.ernet.in}

\author{Ajay Jaiswal}
\affiliation{%
  \institution{Dept.  of Computer Science and Engg.}
  \streetaddress{IIT Kharagpur, India}
  }
\email{ajayjaiswal@iitkgp.ac.in}

\author{Priya Shree}
\affiliation{%
  \institution{Dept.  of Computer Science and Engg.}
  \streetaddress{IIT Kharagpur, India}
  }
\email{priya.shree@iitkgp.ac.in}

\author{Arindam Pal}
\affiliation{%
 \institution{TCS Innovation Labs, India}
}
\email{arindam.pal1@tcs.com}

\author{Animesh Mukherjee}
\affiliation{%
  \institution{Dept.  of Computer Science and Engg.}
  \streetaddress{IIT Kharagpur, India}}
  
\email{animeshm@cse.iitkgp.ernet.in}

\author{Pawan Goyal}
\affiliation{%
  \institution{Dept.  of Computer Science and Engg.}
  \streetaddress{IIT Kharagpur, India}
  }
\email{pawang@cse.iitkgp.ernet.in}

\renewcommand{\shortauthors}{Singh et al.}

\begin{abstract}
This paper explores an interesting new dimension to the challenging problem of predicting long-term scientific impact ($LTSI$) usually measured by the number of citations accumulated by a paper in the long-term. It is well known that early citations (within 1--2 years after publication) acquired by a paper positively affects its $LTSI$. However, there is no work that investigates if the set of authors who bring in these early citations to a paper also affect its $LTSI$. In this paper, we demonstrate for the first time, the impact of these authors whom we call \textit{early citers} (EC) on the $LTSI$ of a paper. Note that this study of the complex dynamics of EC introduces a brand new paradigm in citation behavior analysis. Using a massive computer science bibliographic dataset we identify two distinct categories of EC -- we call those authors who have high overall publication/citation count in the dataset as \textit{influential} and the rest of the authors as \textit{non-influential}. We investigate three characteristic properties of EC and present an extensive analysis of how each category correlates with $LTSI$ in terms of these properties. In contrast to popular perception, we find that influential EC negatively affects $LTSI$ possibly owing to \textit{attention stealing}. To motivate this, we present several representative examples from the dataset. A closer inspection of the collaboration network reveals that this stealing effect is more profound if an EC is nearer to the authors of the paper being investigated. As an intuitive use case, we show that incorporating EC properties in the state-of-the-art supervised citation prediction models leads to high performance margins. At the closing, we present an online portal to visualize EC statistics along with the prediction results for a given query paper. We make all the codes and the processed dataset available in the public domain at our portal:  \url{http://www.cnergres.iitkgp.ac.in/earlyciters/}
\end{abstract}

 \begin{CCSXML}
 <ccs2012>
<concept>
<concept_id>10002951.10003227.10003351</concept_id>
<concept_desc>Information systems~Data mining</concept_desc>
<concept_significance>500</concept_significance>
</concept>
<concept>
<concept_id>10002951.10003227.10003392</concept_id>
<concept_desc>Information systems~Digital libraries and archives</concept_desc>
<concept_significance>500</concept_significance>
</concept>
</ccs2012>
\end{CCSXML}

\ccsdesc[500]{Information systems~Data mining}
\ccsdesc[500]{Information systems~Digital libraries and archives}
\keywords{Long-term scientific impact, citation count, early citers, supervised regression models}
\maketitle

\vspace{-0.5cm}
\section{Introduction}
Success of a research work is estimated by its scientific impact. Quantifying scientific impact through citation counts or metrics~\cite{bergstrom2008eigenfactor,egghe2006theory,garfield1999journal,hirsch2005index} has received much attention in the last two decades. This is primarily owing to the exponential growth in the literature volume requiring the design of efficient impact metrics for policy making concerning with recruitment, promotion and funding of faculty positions, fellowships etc. Although these approaches are quite popular, they appear to be highly debatable~\cite{hirsch2014meaning,labbe2010ike}. Additionally, they fail to take into account the future accomplishments of a researcher/article.
A natural and intriguing question is -- \textit{why should one be concerned about the future accomplishments of a researcher/article?} When an early-career researcher is selected for a tenure-track position, it is an investment. More likely, an organization will largely invest on a researcher who has higher chances of accomplishing more in future. Similarly, to ensure high quality search/recommendation results, search engines can rank recently published articles (low cited) higher than older articles (highly cited), if there is some guarantee that the recent article is going to be popular in the near future. 

Prediction of future citation counts is an extremely challenging task because of the nature and dynamics of citations~\cite{Chakraborty:2014,Singh:2015:RCC:2806416.2806566,yan2012better}. Recent advancement in prediction of future citation counts has led to  the development of complex mathematical and machine learning based models. The existing supervised models have employed several paper, venue and author centric features that can be obtained at the publication time. There are equally many works~\cite{bornmann2013percentile,10.1371/journal.pone.0112520,Wang2013} that leverage citation information generated within 1--2 years after publication to enhance the prediction. Despite this enormous interest, the characteristics of early citations generated immediately after publications have not been dealt with in-depth. In particular, to the best of our knowledge, there is no work that has studied the effect of the early citing authors on the long-term scientific impact ($LTSI$). We would like to stress that here we identify this social process for the first time that introduces a new paradigm in citation behavior analysis. 

The aim of this work is to better understand the complex nature of the \textit{early citers} (EC) and study their influence on $LTSI$. \iffalse Specifically we ask: \textit{how does the early citers impact the long-term scientific impact?} 
To explain this complex nature, the paper introduces \textit{early citers} (EC) for in-depth analysis of early citations.\fi EC represents the set of authors who cite an article early after its publication (within 1--2 years). We investigate three characteristic properties of EC and present an extensive analysis to answer three interesting research questions: %\mycheck{Reformulate?}
\begin{itemize}[noitemsep,nolistsep]
\item Do early citers influence the future citation count of the paper? 
\item How do early citations from influential authors impact the future citation count compared to the non-influential ones?
\item How do citations from co-authors impact the future citation count compared to the others (influential as well as non-influential)?
\end{itemize}

In Section~\ref{sec:expRel}, we present a large-scale empirical study to answer these questions. Motivated by the empirical observations, in Section~\ref{sec:citmodel}, we incorporate the EC features in a popular citation prediction framework proposed by Yan et al.~\cite{yan2012better}. In Section~\ref{sec:analysis}, we discuss the prediction outcomes and show that our extended framework outperforms the original framework by a high margin.
 In particular, we make the following contributions:
\begin{enumerate}[noitemsep,nolistsep]
\item We identify two important categories of EC -- we call those authors that have high publication/citation count in the data as \textit{influential} and the rest of the authors as \textit{non-influential}.
\item We analyze three different characteristic properties of EC.
\item We empirically show that early citations might not be always beneficial; in particular early citations from influential EC negatively correlates with the $LTSI$ of a paper.
\item We build a citation prediction model incorporating the EC features; the prediction outcomes by far outperforms the baseline predictions.  
\item We construct an online portal to present visualization of EC statistics and prediction results for a given query paper. 
\end{enumerate}

\vspace{-0.25cm}
\section{\underline{Early} (Non-)Influential Citers}
\label{sec:early_citers}
The term \textit{early citations} refers to citations accumulated immediately after the publication. In the literature, although, there seems to be no general definition of `early', majority of the works kept it within $\sim$ 2 years after publication~\cite{Singh:2015:RCC:2806416.2806566,Adams2005}. Multiple previous works assert that early citation count helps in better prediction of the $LTSI$~\cite{Chakraborty:2014,bornmann2013percentile,Adams2005}. Although these approaches are interesting, they fail to capture the existence of different types of \textit{early citations} leading to more complex influence patterns on $LTSI$. 

Given a candidate paper $P$ published in the year $T$, we are interested in the citation information generated within $\delta$ year(s) after publication, i.e., within the time interval $[T,T+\delta]$. For example, for $\delta=2$, if an article is published in the year 2000, we look into the citation information generated till 2002. \textit{Early citation
count} $ECC_\delta(P)$ refers to the total number of citations received by the paper $P$ from other articles within $\delta$ years after publication. Note, $ECC_\delta(P)$ quantitatively measures the early popularity of the paper $P$. However, $ECC_\delta(P)$ fails to capture the inherent nature of the individual early citations; for example, there exists no distinction between:
\begin{itemize}[noitemsep,nolistsep]
\item originators (authors, journals etc.) of early citations.
\item good (substantiating) and bad (criticizing) citations.
\item self and non-self citations.
\end{itemize}

\begin{figure}[!tbh]
\vspace{-0.1cm}

 \resizebox{0.7\linewidth}{!}{\includegraphics{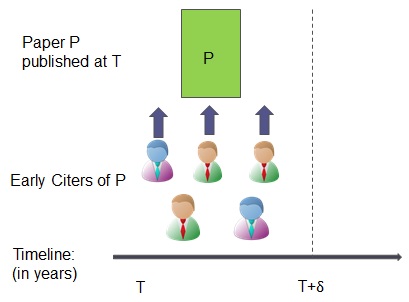}}
 \vspace{-0.3cm}
 \caption{Schematic representation of early citers on a temporal scale. Early citers consist of all authors that cite paper $P$ within  $\delta$  year(s) after its publication. The set of early citers is divided into two subsets, namely, a) influential, and b) non-influential. Influential early citers are represented in purple color (online) whereas non-influential early citers are represented in green color (online).} \label{fig:earlyCiters} 
\vspace{-0.2cm}
\end{figure}

To incorporate some of the above distinctive characteristics in $ECC_\delta(P)$ and to better understand the inherent nature of the individual citations, we present the following three definitions: 

\textbf{Early citers ($EC_\delta(P)$)}: $EC_\delta(P)$ represents the set of authors that cite paper $P$ within $\delta$ years
after its publication. Figure~\ref{fig:earlyCiters} shows schematic representation
of $EC_\delta(P)$ on a temporal scale. Here, $EC_\delta(P)$ consists of all authors that cite paper $P$ within $\delta$ year after its publication. Further, we divide this set into two subsets -- i) \textit{influential}, and ii) \textit{non-influential} early citers.

\noindent\textbf{Influential early citers ($IEC_\delta(P)$)}:  This is a subset of $EC_\delta(P)$ in which each author either has a high publication count or a high citation count or both at the time of citation. Note that, in the current work, we consider top $\sim 5$\% authors as influential early citers, both in terms of publication and citation counts. Empirically
(from dataset described in Section~\ref{sec:dataDes}), we find that top $\sim 5$\% consists of authors who have authored at least 21 publications or acquired atleast 250 citations
or both. In Figure~\ref{fig:earlyCiters}, for paper $P$, $IEC_\delta(P)$ are represented in the purple color.

\noindent\textbf{Non-influential early citers ($NEC_\delta(P)$)}: Early citers that are not influential constitutes the set of non-influential citers, i.e. 
\begin{equation}
NEC_\delta(P)= EC_\delta(P) \setminus IEC_\delta(P)
\end{equation}
As described before, $NEC_\delta$ consists of the remaining $\sim 95$\% of the authors in $EC_\delta(P)$. 
In figure~\ref{fig:earlyCiters}, $NEC_\delta(P)$ authors are represented in green color. 
To study the impact of influential and non-influential EC on citations gained at a later point in time, we define 
long-term scientific impact as:

\noindent\textbf{Long-term scientific impact ($LTSI_\Delta(P)$)}: Given a paper $P$, it represents cumulative citation count of $P$ after $\Delta$ years of its publication. Section~\ref{sec:expRel} demonstrates the effect of influential and non-influential EC on $LTSI$. Next, we describe the dataset we employ for the large scale experimental study and for the extended prediction framework. 

\vspace{-0.25cm}
\section{Dataset Description}
\label{sec:dataDes}
In this paper, we utilize two open source computer science datasets, both crawled from the Microsoft Academic Search (MAS)\footnote{http://academic.research.microsoft.com}. First dataset (bibliographic dataset) was crawled by Chakraborty et al.~\cite{Chakraborty:2014} for a similar prediction work. The dataset consists of bibliographic information of more than 2.4 million papers, such as, the title, the abstract, the keywords, its author(s), the affiliation of the author(s), the year of publication, the publication venue, and the references. Second dataset (citation context dataset) was prepared by Singh et al.~\cite{Singh:2015:RCC:2806416.2806566}. This dataset consists of more than 26 million citation contexts, pre-processed and annotated with the cited and the citing paper information. We combine the above two separately crawled datasets into a single \textit{compiled dataset}.

We filter the compiled dataset by removing papers with incomplete information about the title, the abstract, the venue, the author(s), etc. Since the current study entirely focuses on early citers, we only include papers that consist of at least one citation within $\delta(=2)$ years after publication. We term this dataset as \textit{filtered dataset}. Table~\ref{tab:dataset} outlines the various statistics for both the datasets. For the rest of this paper, we conduct all our experiments on the filtered dataset unless otherwise stated.

\begin{table}[!thb]
\vspace{-0.2cm}
 \caption{General information about the datasets. We combine the two separately crawled datasets -- a) the bibliographic dataset, and b) the citation context dataset into a single compiled dataset. We create the filtered dataset after removing incomplete information from the compiled dataset. Note, the filtered dataset consists of articles that have at least one citation within $\delta(=2)$ years after publication.}\label{dataset}
   \begin{tabular}{|l|l|l|l|}
  \hline
 &\multirow{2}{*}{\parbox[t]{2mm}{\centering Compiled dataset} }& \multirow{2}{*}{\parbox[t]{10mm}{\centering Filtered dataset}}\\
 &&\\\hline
  No. of publications&2,473,147&949,336\\\hline
  No. of authors&1,186,412&535,543\\\hline
  Year range & 1859--2012& 1970--2010\\\hline
 % Avg. no. of authors per paper&2.53&2.69\\\hline
 % Avg. no. of papers per author&5.43&5.28\\\hline
  No. of citation contexts&26,037,804&11,532,780\\\hline
 \end{tabular}
\label{tab:dataset}
 \vspace{-0.5cm}
\end{table}

\vspace{-0.1cm}
\section{Empirical study}
\label{sec:expRel}
In this section, we plan to empirically investigate how the early citers impact the $LTSI$ of a paper. The section begins by introducing
three properties of early citers, namely, the publication count, the citation count and the co-authorship distance. We describe each property in detail and present correlation (using Pearson Correlation) statistics along with representative examples.

\noindent\textbf{General Setting}: Given a candidate paper $P$, we construct a set of early citing papers $C_P$ that cite $P$ within $\delta$ year(s) after publication. For the current study, we keep $\delta = 2$. From the definition presented in section~\ref{sec:early_citers}, $EC_\delta(P)$ consists of all authors that have written papers present in $C_P$.
Next, for each paper $c \in C_P$, we select one representative author among all co-authors based on different selection criterion (described in Sections~\ref{sec:pub}--\ref{sec:coauthor}). More specifically, each selection criterion refers to one distinguishing property of EC. Further, we construct a representative author subset $REC_\delta(P)$ from the selected authors and present correlation statistics of this newly constructed subset with $LTSI$. Note that $REC_\delta(P) \subseteq EC_\delta(P)$. Next, we define the three key properties of EC that assist in distinguishing early citations.

 \vspace{-0.2cm}
\subsection{Publication count}
\label{sec:pub}
\begin{figure}[!tbh]

 \resizebox{0.8\linewidth}{!}{\includegraphics{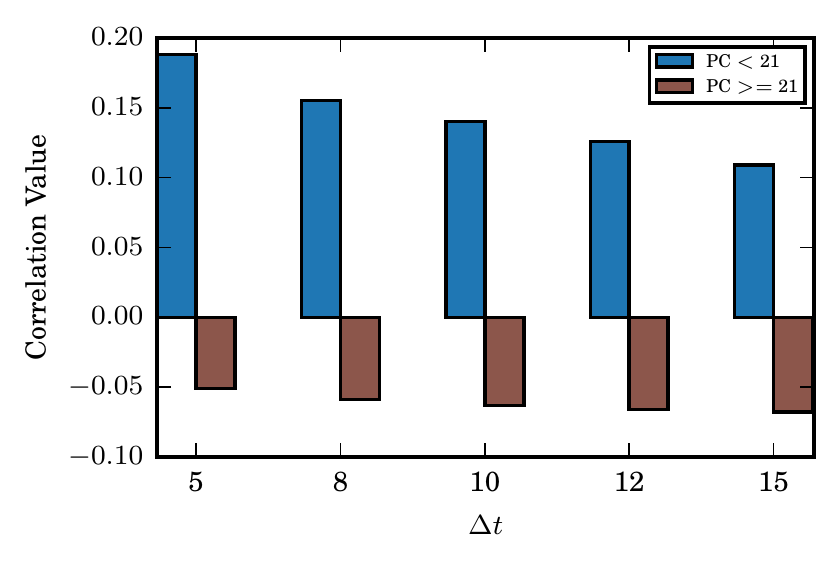}}
 \vspace{-0.5cm}
 \caption{(Color online) Correlation between EC publication count and cumulative citation count at five later time periods after publication, $\Delta t=5,8,10,12,15$. Papers with lower value of $PC(< 21)$ exhibit positive correlation diminishing over the time. Papers with high value of $PC(>= 21)$ show an opposite trend. The overall separation decreases over time.}\label{fig:corrPub}
  \vspace{-0.2cm}
\end{figure}

Publication count of an early citer refers to the number of articles written by her before citing the paper $P$. High publication count denotes high productivity of an early citer. For each paper $c \in C_P$, we select the author with the maximum publication count. The authors so selected constitute the set $REC_\delta(P)$. Note that in our experiments, authors with minimum, average and median publication counts have not shown significant correlations. Further, we aggregate early citers' publication counts ($PC_{P}$) by averaging over the set of selected authors $REC_\delta(P)$. For each paper $P$ present in our dataset, we compute $PC_{P}$ and $P$'s cumulative citation count at five later time periods after publication, $\Delta t=5,8,10,12,15$. We utilize the definitions of influential and non-influential early citers described in section~\ref{sec:early_citers}, i.e., a paper $P$ is cited by a set of influential early citers, if $PC_P >= 21$. Therefore, we split the entire paper set into two subsets: i) papers cited by non-influential EC ($PC_P < 21$), and ii) papers cited by influential EC ($PC_P >= 21$).
Figure~\ref{fig:corrPub} compares  these two subsets correlating $PC$ values with cumulative citation counts at five later time periods. 

 \vspace{-0.3cm}

\noindent\textbf{Observations}: Figure~\ref{fig:corrPub} presents few interesting observations. Papers with lower value of $PC (< 21)$ exhibit positive correlation. However, as $\Delta t$ progresses, this positive correlation starts diminishing. Surprisingly, papers with higher values of $PC (>= 21)$, show negative correlation and this effect becomes more profound as $\Delta t$ progresses. Thus, the overall separation between the two subsets decreases over time. 

This study illustrates the fact that influential EC negatively affect the long-term citations. A plausible  explanation could be that in general, researchers tend to cite works written by influential authors. Therefore, once an influential author cites an article, researchers tend to cite the influential author's paper, instead of the original paper. The attention from the original paper moves to the paper written by the influential citer toward the very beginning of the life-span of the original paper. Therefore, instead of flourishing, the long term citation count of the original paper gets negatively affected. This phenomenon of attention relaying from the less popular article to the more popular article is described as \textit{attention stealing}~\cite{10.1371/journal.pone.0150588}. In case of non-influential EC, the citation count of the candidate paper exhibits a positive correlation with PC. However, with the passage of time, this positive correlation diminishes due to ageing effect associated with paper's life span~\cite{wang2013quantifying}. In case of influential EC, same ageing effect leads to increase in the negative correlation over the passage of time.

Table~\ref{tab:examplePC} shows some specific examples of papers having the same early citation count in the first two years after publication but different PC values. In both cases, the paper having a low PC value receives a much higher citation count in the future.
 
\begin{table}[!thb]
 \vspace{-0.2cm}
 \centering
 \caption{Example paper-pairs having a similar early citation count in the initial two years of publication but different PC values.}\label{dataset}
 \vspace{-0.3cm}
  \resizebox{0.95\textwidth}{!}{\begin{minipage}{\textwidth}
  \begin{tabular}{|c|c|c|c|}
  \hline
\multirow{2}{*}{Paper ID}&\multirow{2}{*}{\parbox[t]{23mm}{\centering Early Citation Count}}& \multirow{2}{*}{\parbox[t]{15mm}{\centering Early citer PC}}&\multirow{2}{*}{\parbox[t]{23mm}{\centering  Later Citation count}}\\
&&&\\\hline
726084&13&18.9&79\\\hline
140790&13&36.5&34\\\hline
\hline
1663998&8&19.17&109\\\hline
150167&8&65&38\\\hline
\end{tabular}
 \end{minipage}}
\label{tab:examplePC}
 \vspace{-0.5cm}
\end{table}

\subsection{Citation count}
\begin{figure}[!tbh]
 \vspace{-0.3cm}
 \resizebox{0.8\linewidth}{!}{\includegraphics{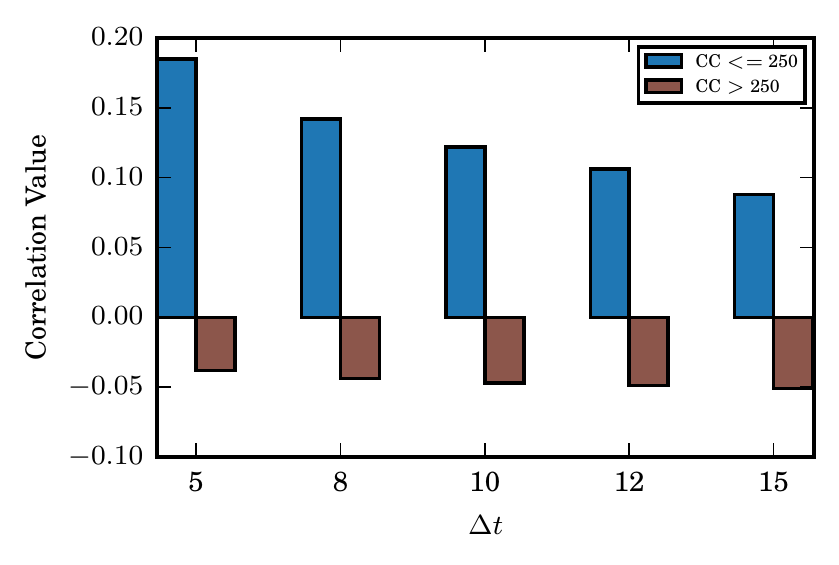}}
 \vspace{-0.5cm}
 \caption{(Color online) Correlation between EC citation count and cumulative citation count at five later time periods after publication, $\Delta t=5,8,10,12,15$. Papers with lower value of $CC(< 250)$ exhibit positive correlation diminishing over the time. Papers with high value of $CC(>= 250)$ show an opposite trend. The overall separation decreases over time.}\label{fig:corrCit}
\end{figure}

Citation count of an early citer refers to the number of citations received by her before citing paper $P$. High citation count denotes higher popularity of the early citer. Again, for each paper $c \in C_P$, we select the author with maximum citation count. Here again, the authors so selected constitute the set $REC_\delta(P)$.
Further, we aggregate early citers' citation counts ($CC_{P}$) by averaging over the set of selected authors $REC_\delta(P)$. For each paper $P$ present in our dataset, we compute $CC_{P}$ and $P$'s cumulative citation count at five later time periods after publication, $\Delta t=5,8,10,12,15$. Similar to previous section, we again split the entire paper set into two subsets: i) papers cited by non-influential EC ($CC_P < 250$), and ii) papers cited by influential EC ($CC_P >= 250$). Figure~\ref{fig:corrCit} compares these two subsets by  correlating $CC$ values with the cumulative citation counts at five later time periods. 

\noindent\textbf{Observations}: Figure~\ref{fig:corrCit} presents similar observations as reported in Figure~\ref{fig:corrPub}. Papers with lower value of $CC(< 250)$ exhibit positive correlation diminishing over the time. Papers with high value of $CC(>= 250)$ show an exactly opposite trend. Here also, the overall separation decreases with time. The results again confirm the existence of \textit{attention stealing}, i.e. a popular citer steals the attention from a newly born paper by citing it. The temporal increase and decrease in correlation values of influential and non-influential early citers respectively relates to the ageing effect as discussed in the previous section.

\begin{table}[!thb]
 \centering
 \caption{Example paper-pairs having a similar early citation count in the initial two years of publication but different CC values.}\label{dataset}
 \vspace{-0.3cm}
  \resizebox{0.95\textwidth}{!}{\begin{minipage}{\textwidth}
  \begin{tabular}{|c|c|c|c|}
  \hline
\multirow{2}{*}{Paper ID}&\multirow{2}{*}{\parbox[t]{23mm}{\centering Early Citation Count}}& \multirow{2}{*}{\parbox[t]{15mm}{\centering Early citer CC}}&\multirow{2}{*}{\parbox[t]{23mm}{\centering  Later Citation count}}\\
&&&\\\hline
2025205&4&124.75&51\\\hline
287142&4&456&13\\\hline
\hline
269672&18&74.45&61\\\hline
1695635&18&623.17&29\\\hline
\end{tabular}
 \end{minipage}}
\label{tab:exampleCC}
 \vspace{-0.4cm}
\end{table}

Table~\ref{tab:exampleCC} shows some specific examples of papers having the same early citation count in the first two years after publication but different CC values. Similar to publication count, here also, we observe that in both the cases, the paper having a low CC value receives a much higher citation count in the future.

\subsection{Co-authorship distance}
\label{sec:coauthor}
We construct a collaboration graph $G(V,E)$ to understand the effect of the co-authorship distance between EC and the authors of candidate paper $P$ on $LTSI$. Here, $V$ is the set of vertices representing authors and an edge $e \in E$ between two authors denotes that they have co-authored at least one article. We define the co-authorship distance (CA) between two authors as \textit{the shortest distance between the two in the co-authorship network}. Again, for each paper $c \in C_P$, we select the author with the lowest $CA$ from the authors of candidate paper $P$. The authors so selected constitutes the set $REC_\delta(P)$ here. Note that in our experiments, authors with highest, average and median co-authorship distance have not shown better correlations. We aggregate the co-authorship distance ($CA_{P}$) by averaging over the set of selected authors $REC_\delta(P)$.
To understand the effect of co-authorship distance on $LTSI$, we divide $CA$ into three buckets:
\begin{itemize}[noitemsep,nolistsep]
\item \textbf{Bucket 1}: $0 \leq CA <1$
\item \textbf{Bucket 2}: $1 \leq CA < 2$
\item \textbf{Bucket 3}: $CA \geq 2$
\end{itemize}

Note, $CA=0$ represents self citations, i.e., one of the early citer is the author of the candidate paper $P$. The authors at $CA= 1$ are the co-authors of the authors in the candidate paper. Hence, \textbf{Bucket 1} mainly consists of authors of the candidate paper itself. \textbf{Bucket 2} mainly consists of the immediate co-authors of the author set of the candidate paper while \textbf{Bucket 3} mainly consists of co-authors of co-authors (distant neighbours) of the author set of the candidate paper.

\begin{figure}[!tbh]
\vspace{-0.2cm}
 \resizebox{0.9\linewidth}{!}{\includegraphics{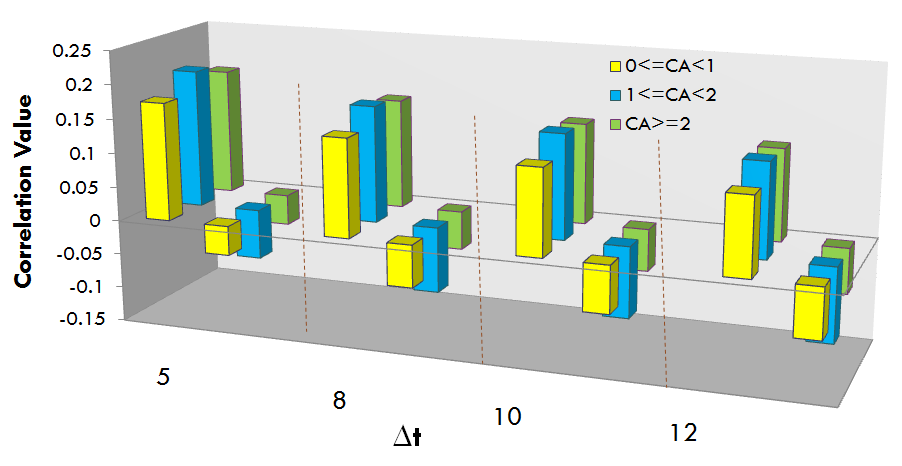}}
 \vspace{-0.3cm}
 \caption{(Color online) Correlation between EC's publication count and cumulative citation count for three co-authorship buckets at four later time periods after publication, $\Delta t=5,8,10,12$. For each time period, first three bars represent correlation for non-influential EC ($PC_P < 21$) whereas the next three bars represent correlation for influential EC ($PC_P >= 21$).  Influential immediate co-authors (\textbf{Bucket 2}) seem to badly affect the citation of the candidate paper $P$ in the long term.}
\label{fig:corrCAPC}
 \vspace{-0.3cm}
\end{figure}

For each bucket, we present correlation statistics of EC's publication count and citation count with $LTSI$. Figure~\ref{fig:corrCAPC} illustrates, for each bucket, correlation between EC's publication count and cumulative citation count at four later time periods after publication, $\Delta t=5,8,10,12$. For each time period, the first three bars represent correlation for non-influential EC ($PC_P < 21$) whereas the next three bars represent correlation for influential EC ($PC_P >= 21$). 

\noindent\textbf{Observations}: For each CA bucket, we observe similar trends as before, influential EC negatively affect the $LTSI$ while non-influential EC affect positively. The most striking observation from this experiment is the effect of immediate co-authors (\textbf{Bucket 2}) on $LTSI$. Even though, both influential or non-influential immediate co-authors maximally correlate with $LTSI$, influential immediate co-authors negatively affect the citation of the candidate paper $P$ in the long term due to intensified \textit{attention stealing} effect. 
\vspace{-0.3cm}

\begin{figure}[!tbh]
\resizebox{0.9\linewidth}{!}{\includegraphics{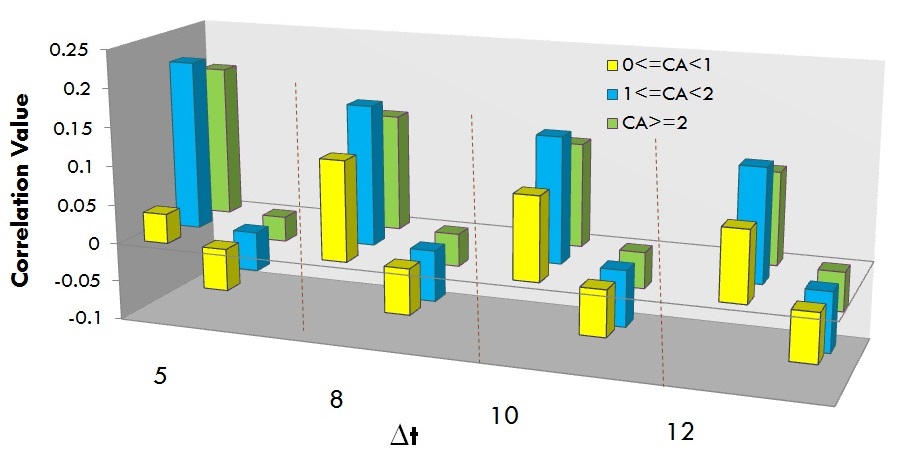}}
\vspace{-0.3cm}
\caption{(Color online) Correlation between EC's citation count and cumulative citation count for three co-authorship buckets at four later time periods after publication, $\Delta t=5,8,10,12$. For each time period, first three bars represent correlation for non-influential EC ($CC_P < 250$) whereas next three bars represent correlation for influential EC ($CC_P >= 250$). Influential immediate co-authors (bucket 2) badly affect the attention of candidate paper $P$ in long term.}\label{fig:corrCACC}
\vspace{-0.2cm}
\end{figure}

Similarly, Figure~\ref{fig:corrCACC} illustrates correlation between EC's citation count and cumulative citation count at four later time periods after publication. For each time period, the first three bars represent correlation for non-influential EC ($CC_P < 250$) whereas the next three bars represent correlation for influential EC ($CC_P >= 250$).

\noindent\textbf{Observations}: In this case, the observations are very similar to the previous case. Motivated by these empirical observations, we incorporate the EC properties in a well recognized citation prediction framework as described in the next section.

\vspace{-0.2cm}
\section{Citation prediction framework}
\label{sec:citmodel}
As an intuitive use case, we extend the long-term citation prediction framework proposed by~\cite{yan2012better} by including the three EC properties discussed in the previous sections. In addition, we also include two citation context based features proposed by Singh et al.~\cite{Singh:2015:RCC:2806416.2806566}. Given a candidate paper, we predict its cumulative citation count at five different time-points ($\Delta t= 3,5,7,9,11$) after publication. Our citation prediction framework employs a set of features that can be computed at the time of publication plus a set of features that can be extracted from the citation information generated within two years after publication (section~\ref{sec:features}). 
We train four predictive models for comparative study, namely, linear regression, Gaussian process regression, classification and regression trees and support vector regression. We discuss each model briefly in Section~\ref{sec:models}. 
We compare our proposed prediction framework with three baselines in Section~\ref{sec:baselines} using evaluation metrics outlined in section~\ref{sec:metrics}.  

\subsection{Feature definition}
\label{sec:features}
As described before, we utilize features available at the time of publication along with the features available within two years after publication. The feature set consists of 20 different features, out of which 14 features are available at the publication time, while the other six features utilize citation information generated within two years after publication. Features\footnote{Some of these features might appear correlated; however, we use all of these in order to have a faithful reproduction of the model proposed in ~\cite{yan2012better}} available at the time of publication are the same as reported in~\cite{yan2012better}. Similarly early citation count and citation context features available after publication are same as reported in~\cite{Singh:2015:RCC:2806416.2806566}. The entire feature set can be divided into seven categories: i) features based on early citer properties, ii) early citation count, iii) features based on paper information, iv) features based on author information, v) features based on venue information, vi) paper recency, and vii) features based on citation context. Given a candidate paper $P$ published in the year $T$, we compute the following features:

\subsubsection{Early citer centric features}
Early citer centric features are computed within two years after the publication. Given a set of early citing papers $C_P$, we compute three features:
\begin{enumerate}
\item \textbf{Publication count (ECPC)}: For each early citing article, we select the author with the maximum publication count. ECPC is computed by averaging this maximum publication count over all the early citing articles.
\item \textbf{Citation count (ECCC)}: Here, for each early citing article, we select the author with the maximum citation count. ECCC is then computed by averaging this maximum citation count over all the early citing articles.
\item \textbf{Co-authorship distance (ECCA)}: Here, we select the author with the minimum co-authorship distance from the authors of the candidate paper $P$. ECCA is computed by averaging this minimum co-authorship distance over all the early citing articles.
\end{enumerate}

\subsubsection{Early citation count (ECC)}
This feature simply includes the citation counts of paper $P$ generated within the first two years after publication. 

\subsubsection{Paper centric features}
\begin{enumerate}
\item \textbf{Novelty (PCN)}: Novelty measures the similarity between paper $P$ and the other publications in the dataset. It is computed by measuring Kullback-Leibler Divergence of an article against all its references. We assume that low similarity means high novelty and more novel article should attract more citations. 
\item \textbf{Topic Rank (PCTR)}: Topics are inferred from the paper title and abstract using unsupervised LDA. Each paper is assigned a topic and further each topic is ranked based on the average citations it has received.
\item \textbf{Diversity (PCD)}: Diversity measures the breadth of an article inferred from its topic distribution. We measure diversity of an article by computing the entropy of the papers's topic distribution (see~\cite{yan2012better} for more details).
\end{enumerate}

\subsubsection{Author centric features}
\begin{enumerate}
\item \textbf{H-Index (ACHI)}: H-index attempts to measure both the productivity and the impact of the published work of a researcher~\cite{hirsch2005index}. Yan et al.~\cite{yan2012better} observed high positive correlation between h-index and average citation counts of publications. 
\item \textbf{Author rank (ACAR)}: Author rank determines the ``fame" of an author. Each author is assigned an author rank based on her current citation count. High rank authors have high citation counts. 
\item \textbf{Past influence of authors (ACPI)}: We measure the past influence of authors in two ways: previous (1) maximum citation counts, and (2) total citation counts. Previous maximum citation count of an author represents the citation count of author's most popular publication. Previous total citation count represents sum of the citation counts of all the author's publications.
\item \textbf{Productivity (ACP)}: The more papers an author has published, the higher average citation counts she could expect. Productivity refers to the total number of articles published by an author. 
\item \textbf{Sociality (ACS)}: A widely connected author is more likely to be cited by her wide variety of co-authors. Sociality, thus, can be computed from the co-authorship network graph employing a formulation in a recursive form as in the PageRank algorithm.
\item \textbf{Authority (ACA)}: A widely cited paper indicates peer acknowledgements, and hence indicates the `authority' of its authors. We compute authority of paper in citation network graph using similar recursive algorithm as proposed for the sociality feature. The paper authority then is transmitted to all its authors.
\item \textbf{Versatility (ACV)}: Versatility represents the topical breadth of an author. We measure the versatility of an author by computing the entropy of the author's topic distribution. Higher versatility implies large volumes of audience from various research fields.
\end{enumerate}

\subsubsection{Venue centric features}
\begin{enumerate}
\item \textbf{Venue rank (VCVR)}: The reputation of a venue relates to the volume of citations it receives. Similar to author rank, we rank venues based on its current citation count. High rank venues have high citation counts. 
\item \textbf{Venue centrality (VCVC)}: We create a venue connective graph $G(V,E)$ where $V$ denotes the set of venues and the edges $e \in E$ denote the citing-cited relationships between venues. The in-degrees measure how many times a venue is cited by papers from other venues. Finally, venue centrality can be measured using a PageRank algorithm.
\item \textbf{Past influence of venues (VCPI)}: Past influence of a venue is computed similar to the past influence of authors. As in the case of authors, we measure the past influence of venues in two ways: previous (1) maximum influence of venues, and (2) total influence of venues. 
\end{enumerate}

\subsubsection{Recency (PR)}
Recency describes the temporal proximity of an article. It measures the age of a published article. The longer an article is published, the more citations it may receive.

\subsubsection{Citation context centric features}
\begin{enumerate}
\item \textbf{Average countX (CCAC)}: A high value of countX implies that the cited paper is referred multiple times by the citer paper in different sections of its text. Thus, cited paper might be quite relevant for citing paper. Singh et al.~\cite{Singh:2015:RCC:2806416.2806566} argued that highly cited papers are cited more number of times in a single text.
\item \textbf{Average citeWords (CCAW)}: Similar to countX, a high value of citeWords implies that the cited paper has been discussed in more details by the citer paper and therefore, cited paper might be quite relevant for the citing paper.
\end{enumerate}

\subsection{Predictive models}
\label{sec:models}

In this section, we describe four regression models. Each model is trained on features described in previous section. All models are trained using available implementations from the Weka toolkit~\cite{hall2009weka}. 

\subsubsection{Linear regression (LR)}	
Linear regression is an approach to model the relationship between the dependent variable $Y$ and one or more independent (explanatory) variables $X$. It attempts to model this relationship by fitting a linear equation to observed data. A linear regression line has an equation of the form:
\begin{equation}
Y = wX^T +b,
\end{equation}
 where $Y$ is the dependent variable, $X^T$ is a vector of explanatory variables, $w$ is a vector of weights (parameters) of the linear regression and $b$ represents the error. In the current work, we consider publication's predicted citation count to be the dependent variable and features (described in Section~\ref{sec:features}) are considered to be the explanatory variables.

\subsubsection{Gaussian process regression (GPR)}
Due to the complex nature of the long-term citation impact estimation, it might well be the case that the dependent variable is a non-linear function of all the features used to represent the data. Gaussian processes~\cite{rasmussen2006gaussian} provide formulations by which the prior information about the regression parameters can be easily encoded.  This property makes them convenient for our problem formulation. Given a vector of input features $X$, the predicted citation counts $C(d)$ of the document $d$ is:
\begin{equation}
C(d) = K(X,X^T)[K(X^T,X^T) + \sigma^2I]^{-1} C(d^T) ,
\end{equation}
where $X^T$ is a matrix of feature vectors of the training set, $K$ is a kernel function, $I$ is the identity matrix, $\sigma$ is the noise parameter and $C(d^T)$ is the vector of citation counts of the training set. Note, in our experiments, we keep $\sigma= 0.5$.

\subsubsection{Classification and regression trees (CART)}
Classification and regression trees~\cite{breiman1984classification} are obtained by recursively partitioning the training data space and fitting a simple prediction model within each partition. As a result, the partitioning can be represented graphically as a decision tree.	Regression trees are built for dependent variables (citation count in the present context) that take continuous or ordered discrete values, with prediction error typically measured by the squared difference between the observed and predicted values. 

\subsubsection{Support vector regression (SVR)}
Support vector regression~\cite{smola1997support} are derived from statistical learning theory and they work by solving a constrained quadratic problem where the convex objective function for minimization is given by the combination of a loss function with a regularization term. Support vector regression is the most common application form of SVMs. In the current study, we employ LIBSVM\footnote{http://www.csie.ntu.edu.tw/$\sim$cjlin/libsvm/} with default parameter settings. The best results were obtained for the linear kernel.

\subsection{Baselines}
\label{sec:baselines}

\subsubsection{Baseline I}
\label{sec:baselineI}
The first baseline~\cite{yan2012better} is similar to our model except that it does not include any information generated after the publication. It includes paper, author and venue centric features along with recency.  

\subsubsection{Baseline II}
\label{sec:baselineII}
The second baseline is similar to Baseline I plus one more feature -- early citation counts.
Chakraborty et al.~\cite{Chakraborty:2014} showed that inclusion of early citation counts enhances prediction accuracies mostly for the higher values of $\Delta t$.

\subsubsection{Baseline III}
\label{sec:baselineIII}
In the third baseline, we include citation context centric features introduced by Singh et al.~\cite{Singh:2015:RCC:2806416.2806566} to Baseline II. Thus, baseline III consists of paper, author, venue and citation context centric features along with recency and early citation count.

\subsection{Evaluation metrics}
\label{sec:metrics}
%We use two evaluation metrics to compare our model with baselines (described in Section~\ref{sec:baselines}). 

\subsubsection{Coefficient of determination ($R^2$)}
Coefficient of determination ($R^2$)~\cite{cameron1997r} measures how well the data fits a statistical model of future outcome prediction. It determines the variability introduced by the statistical model. Let $d$ be the document in the test document set $D$, we compute $R^2$ as:

\begin{equation}
R^2 = \frac{\sum_{d\epsilon{D}}{{(C_p(d) - C_a(D) ) }^2} }{\sum_{d\epsilon{D}}{{(C_a(d) - C_a(D) ) }^2}}
\end{equation}

Here, $C_p(d)$ denotes the predicted citation count for document $d$. $C_{a}(D)$ denotes the mean of observed citation counts for the documents in $D$. $C_{a}(d)$ denotes actual citation count for document $d$. $R^2$ values range from $0$ to $1$. A larger value indicates better performance.

\subsubsection{Pearson correlation coefficient ($\rho$)}
Pearson correlation coefficient ($\rho$)~\cite{lee1988thirteen} measures the degree of linear dependence between two variables. Let $d$ be the document in the test document set $D$, we compute $\rho$ as:
\begin{equation}\small
\rho =\frac{\sum_{d\epsilon{D}}(C_p(d) - C_p(D))(C_a(d) - C_a(D))}{\sqrt{\sum_{d\epsilon{D}}(C_p(d) - C_p(D))^2} \sqrt{\sum_{d\epsilon{D}}((C_a(d) - C_a(D))^2}}
\end{equation}
Here, $C_p(d)$ and $C_a(d)$ represents predicted citation count and actual citation count of test document $d$ respectively. $C_p(D)$ and $C_a(D)$ represent mean of the predicted and the observed citation counts for the documents in $D$. $\rho$ ranges 
from -1 to 1, where $\rho=1$ corresponds to a total positive correlation, $0$ corresponds to no correlation, and $-1$ corresponds to total 
negative correlation. A larger value indicates better performance.

\section{Prediction Analysis}
\label{sec:analysis}

\subsection{Experimental setup}
Our experimental setup bears a close resemblance to~\cite{yan2012better}. We
randomly select 10,000 training sample papers published in and before the year 1995. We opted for a small sample size because of associated computational complexities. Since, our prediction framework utilizes information generated within first two years after
publication, we perform prediction task from 1998 -- 2010. The reason behind choosing 1998 as the start year is to counter information leakage due to the training papers published at 1995 since prediction framework utilizes early citation data till 1997 for papers published in the year 1995. To evaluate, we select three random sets of 10,000 sample papers (published between 1998 -- 2010). Note that for $\Delta t =11$, we can only consider papers published between 1998 -- 1999, for $\Delta t =9$, we can consider papers published between 1998 -- 2001 and so on. Given a candidate paper, we predict its cumulative citation count at five different time-points after publication, $\Delta t=3,5,7,9,11$. For example, given a candidate paper $P$ published in 1998, $\Delta t$ = 3 represents
prediction at 2001, $\Delta t$ = 5 represents prediction at 2003 and so on. In the next section, we present a comprehensive analysis of our proposed framework.

\subsection{Prediction results}

\subsubsection{Comparison between predictive models}
\label{subsec:predictive_model}
\noindent\textbf{Our model}: To begin with, we incorporate all features described in section~\ref{sec:features} for the prediction task (includes early citer centric, paper centric, author centric, venue centric, citation context centric features plus early citation count and recency features). However, we observe marginal performance gain in all models after removing the citation context based features. Therefore, it was decided that the best framework (hereafter `\textit{our model}') for this prediction task would consist of all features except the citation context based features. Table~\ref{tab:models} compares the four predictive models (LR, GPR, CART and SVR) at five different time-points after publication, $\Delta t=3,5,7,9,11$. Overall, SVR achieves the best performance, while GPR seems to have the worst performance. As expected, in all the models, the performance diminishes as $\Delta t$ increases.

\begin{table}[!tbh]
\caption{Performance comparison among the four predictive models -- LR, GPR, CART and SVR. Two evaluation metrics $R^2$and $\rho$  are used. A high value of  $R^2$and $\rho$ represent an efficient prediction. Prediction is performed over five time periods, $\Delta t=3,5,7,9,11$.}
\label{tab:models}
\resizebox{0.8\textwidth}{!}{\begin{minipage}{\textwidth}
\begin{tabular}{|c|c|c|c|c|c|c|c|c|c|c|}
\hline
\multirow{2}{*}{Model}& \multicolumn{2}{c|}{$\Delta T = 3$} & \multicolumn{2}{c|}{$\Delta T = 5$} & \multicolumn{2}{c|}{$\Delta T = 7$} & \multicolumn{2}{c|}{$\Delta T = 9$}  & \multicolumn{2}{c|}{$\Delta T = 11$} \\ \cline{2-11} 
& $\rho$ & $R^2$ & $\rho$ & $R^2$& $\rho$ & $R^2$& $\rho$ & $R^2$ &$\rho$ & $R^2$ \\ \hline
LR         & 0.95 &0.82&0.91 &0.79 & 0.84 & 0.74 &0.81 &0.68  &0.75 & 0.61 \\ \hline
GPR      & 0.83 &0.57&0.80 &0.55 & 0.71 & 0.48 &0.66  & 0.47 & 0.64 & 0.30 \\ \hline
CART     &0.95 &0.73&0.87 &0.68 & 0.79 & 0.62 &0.75  &  0.55 &0.63 & 0.43  \\ \hline
SVR      &\textbf{ 0.97}         &   \textbf{0.84} &   \textbf{0.91} &     \textbf{0.82} &     \textbf{0.88}
 & \textbf{0.76}   & \textbf{0.82}   & \textbf{0.69}  &   \textbf{0.76} &   \textbf{0.65}  \\ \hline
\end{tabular}
\end{minipage}}
\vspace{-0.25cm}
\end{table}

\subsubsection{Comparison with the baseline models}
\begin{figure*}[!tbh]
\resizebox{0.75\linewidth}{!}{\includegraphics{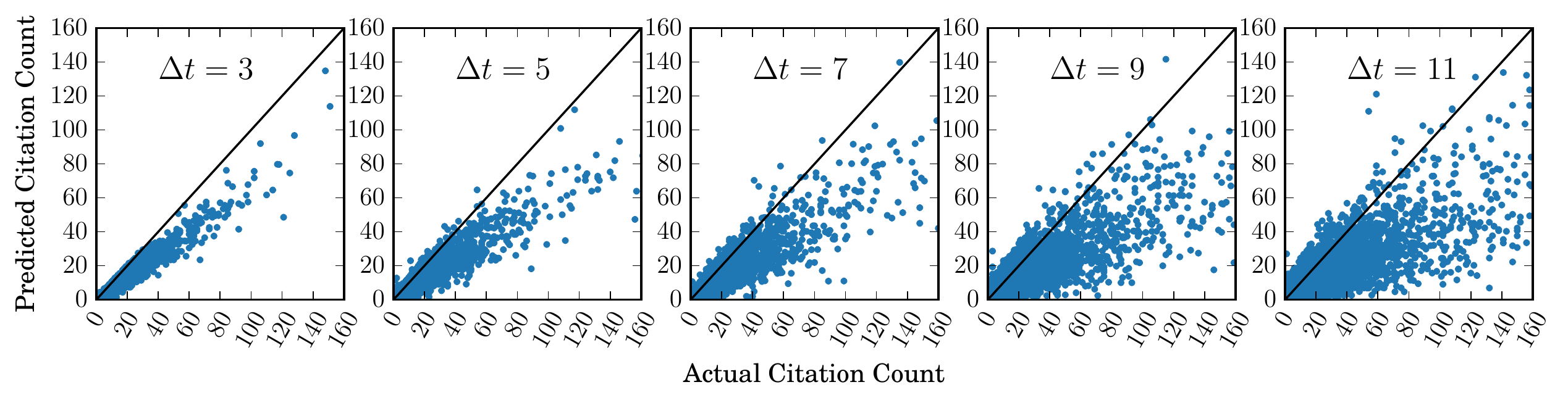}}
\vspace{-0.3cm}
\caption{Change in prediction results over five time-periods. Scatter plots showing correlation between SVR predictions with real citation count values at $\Delta t=3,5,7,9,11$. The black color line represents $y = x$ line passing through origin. Our model performs best for $\Delta T$ = 3 with majority of the points on $y = x$ line. It performs worst for $\Delta T$ = 11 with high divergence from the line. Our model under estimates $LTSI$ as majority of the points lie below the line. However, this prediction is considerably better than all the other baselines.}\label{fig:scatter}
\vspace{-0.4cm}
\end{figure*}

Next, we compare the performance of the three baselines (described in section~\ref{sec:baselines}) with our model. 
Due to high performance gain discussed in the previous section, we use SVR for modeling the three baselines as well as our model. 
Table~\ref{tab:baseline} compares Baseline I, Baseline II and Baseline III with our model. Prediction is made over five time periods, $\Delta t=3,5,7,9,11$. Each cell represents mean and standard deviation (in parenthesis) of the metric values for the three random samples.
Even though, as highlighted, our model by far outperform all three baselines at each time period for both metrics, it slightly under estimates LTSI (see Figure \ref{fig:scatter}).

\begin{table*}[!tbh]
\caption{Performance comparison among Baseline I, Baseline II, Baseline III and our model. Two evaluation metrics  $\rho$ and $R^2$ are used. A high value of  both metrics represent an efficient model. Prediction is made over five time periods, $\Delta t=3,5,7,9,11$. Each cell represents mean and standard deviation (in parenthesis) of the metric values for three random samples. Bold numbers in the table indicate the best performing model for a given time period. Our model by far outperforms all three baselines at each time period for both metrics.}\label{tab:baseline}
\vspace{-0.3cm}
\begin{minipage}{0.9\textwidth}
\begin{tabular}{|c|c|c|c|c|c|c|c|c|}
\hline
\multirow{2}{*}{$\Delta t$} & \multicolumn{2}{c|}{Baseline I} & \multicolumn{2}{c|}{Baseline II} & \multicolumn{2}{c|}{Baseline III} &  \multicolumn{2}{c|}{Our model} \\ \cline{2-9} 
& $\rho$ & $R^2$ & $\rho$ & $R^2$& $\rho$ & $R^2$& $\rho$ & $R^2$ \\ \hline
3&0.793 (0.003)&0.654 (0.019)&0.856 (0.021)&0.724 (0.001)&0.895 (0.012)&0.769 (0.017)&\textbf{0.971} (0.002)&\textbf{0.841} (0.001)\\ \hline
5&0.745 (0.021)&0.644 (0.006)&0.792 (0.007)&0.699 (0.012)&0.814 (0.019)&0.788 (0.001)&\textbf{0.915} (0.015)&\textbf{0.819} (0.019)\\\hline
7&0.691 (0.016)&0.593 (0.003)&0.752 (0.004)&0.688 (0.019)&0.754 (0.023)&0.690 (0.026)&\textbf{0.877} (0.007)&\textbf{0.765} (0.013)\\\hline
9&0.543 (0.008)&0.588 (0.015)&0.646 (0.009)&0.639 (0.002)&0.684 (0.002)&0.643 (0.001)&\textbf{0.819} (0.003)&\textbf{0.687} (0.021)\\\hline
11&0.591 (0.015)&0.544 (0.002)&0.633 (0.010)&0.542 (0.006)&0.675 (0.008)&0.582 (0.021)&\textbf{0.758} (0.005)&\textbf{0.651} (0.016)\\\hline
\end{tabular}
\end{minipage}
\vspace{-0.2cm}
\end{table*}

\subsubsection{Effect of different early time periods}
So far, we have performed experiments for a fixed early time period ($\delta=2$). In this section, we experiment with $\delta=1,2,3$ for estimating the early citer features\footnote{Note that the early citation count however is obtained using  $\delta=2$ as suggested in the literature.}. Table~\ref{tab:early_time_period} compares the prediction results for the SVR model using three different values of $\delta$. The table presents an interesting finding that increasing the value of $\delta$ does not always improve prediction accuracy. $R^2$ values at $\delta = 2$ always outperform $\delta = 1,3$ in the later time points.

\begin{table}[!tbh]
\centering
\caption{Performance of the model assuming different values of $\delta$. 
Prediction is made over three early time periods, $\delta =1,2,3$, and at three later time points, $\Delta t=5,7,9$. Best results are obtained at $\delta$ = 2. The added information does not always improve prediction accuracy.}
\vspace{-0.4cm}
\label{tab:early_time_period}
\begin{tabular}{|c|c|c|c|c|c|c|}
\hline
\multirow{2}{*}{$\Delta T$} & \multicolumn{2}{c|}{$\delta = 1$} & \multicolumn{2}{c|}{$\delta = 2$} & \multicolumn{2}{c|}{$\delta = 3$} \\ \cline{2-7} 
    &$\rho$         & $R^2$       & $\rho$         & $R^2$& $\rho$         & $R^2$\\ \hline
5   & 0.882 & 0.68 & \textbf{0.915} & \textbf{0.82} & 0.911  & 0.76\\ \hline
7   & 0.841& 0.61 & 0.877 & \textbf{0.77}&  \textbf{0.884} & 0.72\\ \hline
9   & 0.765 &  0.58& 0.819& \textbf{0.69}  & \textbf{0.822}  &0.64 \\ \hline
\end{tabular}
\vspace{-0.25cm}
\end{table}

\subsection{Feature analysis}
We now study how the various features correlate with the actual
citation counts. As described in Section~\ref{subsec:predictive_model}, our model is trained on 18 features out of 20 features (described in Section~\ref{sec:features}); therefore, we perform feature analysis for 18 features. We train SVR with individual features and rank them based on Pearson's correlation values of each feature with the actual citation count for $\Delta t=3$ years after publication in descending order. Table~\ref{tab:featureimportance} reports ranked list of features at $\Delta t=3$. We can observe from the table that the first six in the rank list consists of all the three EC features, indicating importance of the EC features. As expected, early citation count is the most distinctive feature.

\begin{table}[!tbh]
\centering
\caption{Ranked list of features based on Pearson's correlation values between the predicted citation count and the actual citation count for $\Delta t=3$ years after publication. Each SVR model is trained with individual feature.}
\label{tab:featureimportance}
\begin{tabular}{|l|c|l|c|l|c|l|c|}
\hline
1 & ECC  & 6 & ECCA & 11 & ACAR & 16 & PCN \\\hline
2 & ECCC & 7 & ACHI & 12 & ACP & 17 & ACV \\\hline
3 & ECPC & 8 & VCVR & 13 & PCTR & 18 & VCVC \\\hline
4 & VCPI & 9 & ACS & 14 & PR & \multicolumn{2}{c}{} \\\cline{1-6}
5 & ACPI & 10 & PCD& 15 & ACA & \multicolumn{2}{c}{} \\\cline{1-6}
\end{tabular}
\vspace{-0.25cm}
\end{table}

\begin{figure}[!tbh]
\resizebox{0.85\linewidth}{!}{\includegraphics{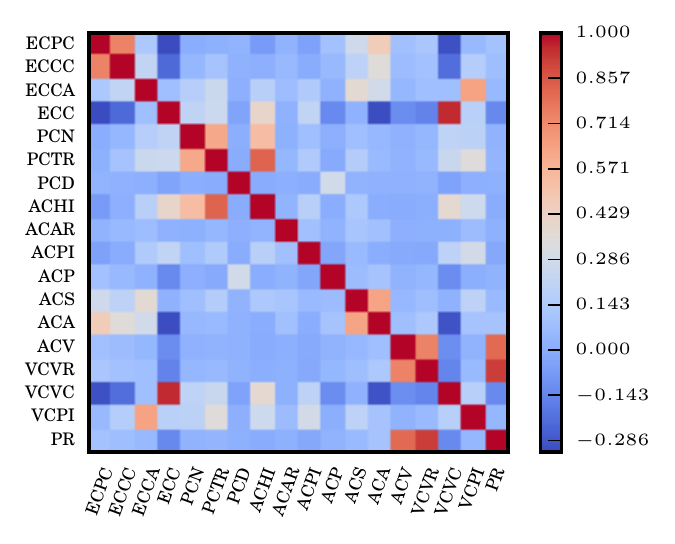}}
\vspace{-0.3cm}
\caption{(Color online) Cross correlation between features: Red color represents highly correlated features (=1). Blue represents uncorrelated to weakly negatively correlated features. Diagonal entries have maximum correlation (self) values = $1$. }\label{fig:feature_correlation}
\vspace{-0.6cm}
\end{figure}

Figure~\ref{fig:feature_correlation} presents cross-correlation between features. Diagonal entries have maximum positive correlation (self) values = 1. Overall, features seem to be not much correlated with each other except a few cases. Interestingly, we observe that the EC features negatively correlate with the early citation count feature, the two being very distinct sources of information. Thus, including the EC features enhances the prediction performance significantly over and above the early citation count feature.

\vspace{-0.2cm}

\begin{figure}[!tbh]
\resizebox{0.85\linewidth}{!}
{\includegraphics{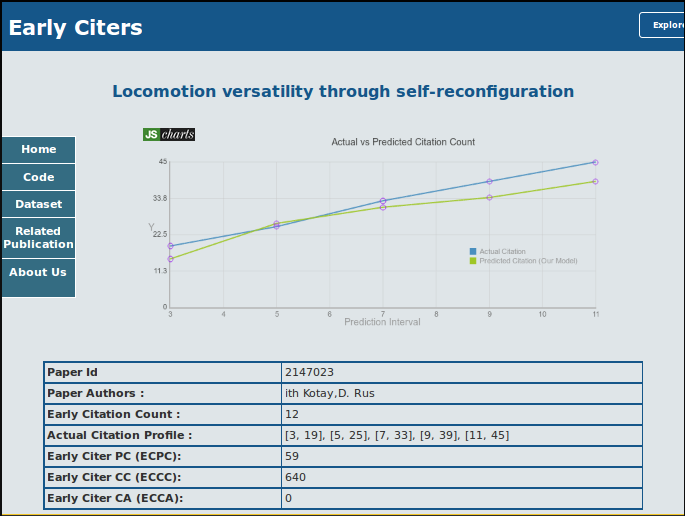}}
\caption{(Color online) Snapshot of online portal: For input candidate paper, the portal presents visualization of prediction results along with EC statistics. It compares SVR predictions with real values  at $\Delta t = 3,5,7,9,11$ years after publication.}
\vspace{-0.5cm}
\end{figure}

\section{Online portal}
\label{sec:onlinePortal}
We have also built an online portal to showcase the different results from our current work. Given a query paper present in our dataset, the portal displays different statistics related to the paper; in particular, each query result is accompanied by the statistics of the EC properties and other paper details. In addition, the portal also presents with a visualization comparing the actual and the predicted citation count of the paper.  The current system is hosted on our research group server and can be accessed at \url{http://www.cnergres.iitkgp.ac.in/earlyciters/}.

\vspace{-0.2cm}
\section{Related Work}
\label{sec:relWork}
In recent years, several researchers have investigated the problem of $LTSI$~\cite{Chakraborty:2014,Singh:2015:RCC:2806416.2806566,wang2013quantifying,yan2012better}.
While some works propose complex mathematical models~\cite{mingers2008exploring,stegehuis2015predicting,wang2013quantifying,Wang2013,Wang20094273,DBLP:conf/ijcai/XiaoYLJWYCZ16} incorporating ageing assumptions, majority of
the works focused on supervised machine learning models. Moreover, there are few recent works~\cite{bornmann2013percentile,Wang2013} that present an empirical analysis of the correlation between short-term and long-term citation counts. Interestingly, Stern~\cite{10.1371/journal.pone.0112520} reports that shortly after the appearance of a publication the combined use of early citations and impact factors yields a better prediction of the $LTSI$ of the publication than the use of early citations only. Recently, Didegah et al.~\cite{didegah2013factors} presented an overview of the literature on predicting $LTSI$. 

\noindent\textbf{Mathematical models}: The use of early citations to predict $LTSI$ has been studied in various papers using mathematical models. Wang et al.~\cite{Wang2013} and Mingers et al.~\cite{mingers2008exploring} proposed models that described how publications accumulate citations over the time. Stegehuis et al.~\cite{stegehuis2015predicting} employed two predictor models (journal impact factor and early paper citations) to predict a probability distribution for the future citation count of a publication. They only considered accumulated citations within one year after publication. This is in contrast to the approach proposed by Wang et al.~\cite{wang2013quantifying} where they allow predictions to be made fairly soon after the appearance of a publication. They propose three fundamental citation driving mechanisms -- a) preferential attachment, b) ageing and novelty, and c) importance of a discovery. Their proposed model collapses the citation histories of papers from different journals and disciplines into a single curve indicating that all papers tend to follow the same universal temporal pattern. More recent work by Xiao et al.~\cite{DBLP:conf/ijcai/XiaoYLJWYCZ16} explored paper-specific covariates and a point process model to account for the ageing effect and triggering role of recent citations.

\noindent\textbf{Machine learning models}: Among machine learning (ML) based prediction models, majority of the works have utilized support vector regression (SVR)~\cite{Chakraborty:2014,Singh:2015:RCC:2806416.2806566}, classification and regression tree (CART)~\cite{callaham2002journal,yan2011citation} and linear and multiple regression models~\cite{kulkarni2007characteristics, lokker2008prediction}. Among ML models, we categorize works
into three types based on the temporal availability of features -- (a) features available
at the time of publication~\cite{callaham2002journal,Lawrence,kulkarni2007characteristics,livne2013predicting,yan2012better}, 
(b) features available after publication~\cite{brody2006earlier}, and c) combination of (a)
and (b)~\cite{Chakraborty:2014,Singh:2015:RCC:2806416.2806566}. 
Callaham et al.~\cite{callaham2002journal} used features like journal impact factor, research design, number of subjects, rated subjectivity for scientific quality, news-worthiness etc. Further, they train decision trees to predict citation counts of 204 publications from emergency medicine specialty meeting. Livne et al.~\cite{livne2013predicting} used five group of features -- authors, institutions, venue, references network and content similarity to train an SVR model. Similarly, Kulkarni et al.~\cite{kulkarni2007characteristics} also used information present at the publication time. They train linear regression to predict citation count for five year ahead window using 328 medical articles. Yan et al.~\cite{yan2012better} introduced features covering venue prestige, content novelty and diversity, and authors' influence and activity. Another work used data generated after the publication to predict citation count~\cite{brody2006earlier}. In this study, the downloaded data within the first six months after publication was used as a predictive feature. 
Chakraborty et al.~\cite{Chakraborty:2014} claimed that stratified learning approach leads to higher prediction accuracy. They proposed a two-stage prediction model that consumes information present at the publication time as well as citation information generated within the first two years after publication. Singh et al.~\cite{Singh:2015:RCC:2806416.2806566} proposed extension to previous work~\cite{Chakraborty:2014} by including crowdsource based textual features like countX and citeWords.

\section{Conclusion and Future Work}
\label{sec:conFuture}
This paper has investigated influence of early citers (EC) on long-term scientific impact. 
We have been successfully able to provide empirical evidence that early citers play a significant role in determining the long-term scientific impact. 
More specifically, we find that influential EC have a negative impact while non-influential EC have a positive impact on a paper's $LTSI$. We have provided further evidence that the
negative impact is more intense when EC is closer to the authors of the candidate article in the collaboration network. Drawing from these observations, we incorporate the EC properties in a state-of-the-art supervised prediction model obtaining high performance gains. We believe that the identification of this social process actually leads to a new paradigm in citation behavior analysis. 

In future, we believe that our work can be easily generalized for other scientific research fields. This study is the first step towards enhancing our understanding of influence of EC. To further our research we plan to analyze effects of EC in the patent datasets as well. Future work will concentrate on mathematical modeling of EC influence. 

%In conclusion, this research has raised many questions in need of further investigation. \mycheck{Not sure if we can / should say the last line.}
\vspace{-0.2cm}
\bibliographystyle{ACM-Reference-Format}
\bibliography{sigproc}

\end{document}